\documentclass[amsmath,notitlepage,preprintnumbers,amssymb,nofootinbib,superscriptaddress]{revtex4-1}

\usepackage[left=1in,right=1in,top=1in,bottom=1in]{geometry}

\usepackage{graphicx}
\usepackage{float}
\usepackage{subcaption}

\usepackage{amsmath}

\usepackage[compact]{titlesec}  
\titlespacing{\section}{10pt}{10pt}{10pt}
\titlespacing{\subsection}{5pt}{5pt}{5pt}
\titlespacing{\subsubsection}{5pt}{5pt}{5pt}

\labelformat{section}{\thesection}
\labelformat{subsection}{\thesubsection}
\labelformat{subsubsection}{\thesubsubsection}

\usepackage{hyperref}

\titleformat{\section}
  {\normalfont\fontsize{12}{15}\bfseries}{\thesection}{1em}{}

\titleformat{\subsection}
  {\normalfont\fontsize{11}{12}\bfseries}{\thesubsection}{1em}{}

\titleformat{\subsubsection}
  {\normalfont\fontsize{11}{12}\bfseries}{\thesubsubsection}{1em}{}
  
\renewcommand{\thesection}{\arabic{section}}
\renewcommand{\thesubsection}{\thesection.\arabic{subsection}}
\renewcommand{\thesubsubsection}{\thesubsection.\arabic{subsubsection}}

\renewcommand{\thetable}{\arabic{table}}

\usepackage{color}
\definecolor{violet}{rgb}{0.5,0,1}
\definecolor{orange}{rgb}{1,0.5,0}
\definecolor{teal}{rgb}{0,0.5,0.5}

\usepackage[normalem]{ulem} 

\usepackage{enumitem}

\usepackage{xspace}

\usepackage{caption}
\newcommand{\unit}[1]{\ensuremath{\mathrm{\,#1}}\xspace}

\newcommand{\e}{\unit{e^{-}}}

\begin{document}


\title{\boldmath DarkNESS: developing a skipper-CCD instrument to search for Dark Matter from Low Earth Orbit}

\author{Phoenix Alpine}
\email{alpine2@illinois.edu}
\affiliation{University of Illinois, Urbana-Champaign, IL, 61801, USA}

\author{Samriddhi Bhatia}
\affiliation{University of Illinois, Urbana-Champaign, IL, 61801, USA}

\author{Fernando Chierchie}
\affiliation{Instituto de Inv. en Ing. El\'ectrica ``Alfredo C. Desages'' (IEEE) CONICET, Bah\'ia Blanca, Argentina}

\author{Alex Drlica-Wagner}
\affiliation{Fermi National Accelerator Laboratory, Batavia, IL 60510, USA}
\affiliation{Department of Astronomy and Astrophysics, University of Chicago, Chicago, IL 60637, USA}

\author{Rouven Essig}
\affiliation{ 
C.N.~Yang Institute for Theoretical Physics, Stony Brook University, Stony Brook, NY 11794, USA}

\author{Juan Estrada}
\affiliation{Fermi National Accelerator Laboratory, Batavia, IL 60510, USA}
\affiliation{Department of Astronomy and Astrophysics, University of Chicago, Chicago, IL 60637, USA}

\author{Erez Etzion}
\affiliation{School of Physics and Astronomy, Tel Aviv University, Tel Aviv, 69978, Israel}

\author{Roni Harnik}
\affiliation{Fermi National Accelerator Laboratory, Batavia, IL 60510, USA}

\author{Michael Lembeck}
\affiliation{University of Illinois, Urbana-Champaign, IL, 61801, USA}

\author{Nathan Saffold}
\email{nsaffold@fnal.gov}
\affiliation{Fermi National Accelerator Laboratory, Batavia, IL 60510, USA}
\affiliation{Kavli Institute for Cosmological Physics, University of Chicago, Chicago, IL 60637, USA}

\author{Sho Uemura}
\affiliation{Fermi National Accelerator Laboratory, Batavia, IL 60510, USA}


\preprint{FERMILAB-FN-1265-PPD}


\begin{abstract}
The DarkNESS (Dark Matter Nano-satellite Equipped with Skipper Sensors) mission aims to deploy a skipper-CCD CubeSat Observatory to search for dark matter (DM) from Low Earth Orbit.
This mission will employ novel skipper-CCDs to investigate $\mathcal{O}\mathrm{(keV)}$ X-rays from decaying DM, as well as electron recoils from strongly-interacting sub-GeV DM.
The DarkNESS mission will be the first space deployment of skipper-CCDs, and the DarkNESS team is developing a skipper-CCD instrument that is compatible with the CubeSat platform. DarkNESS has  
recently progressed from laboratory validation to a Critical Design Review (CDR) phase, with
a launch opportunity anticipated in late 2025.
The implementation of the DarkNESS skipper-CCD payload on the CubeSat platform will pave the way for future demonstrators of space-based imagers for X-ray and single-electron counting applications.

\end{abstract}

\maketitle


\section{Introduction}
\label{sec:overview}

One of the key pursuits of modern physics is to uncover the nature of dark matter (DM), which dominates the cosmic mass budget but has eluded detection~\cite{Planck_2020,Arbey_2021}. While searches for Weakly Interacting Massive Particles (WIMPs) have been the focus of DM experiments in recent decades~\cite{Akerib_2022_snowmass}, persistent non-detection demands a broader exploration of potential DM candidates. Most DM direct detection experiments are conducted underground to mitigate cosmogenic backgrounds; however, there are potential signatures of DM that are inaccessible to underground experiments due to attenuation in the Earth’s atmosphere and crust.

Space-based observations are paramount to detect astrophysical signals that do not penetrate the Earth's atmosphere. 
Furthermore, the development of the CubeSat platform has enabled the deployment of compact instrumentation in space with the capability to significantly impact focused scientific goals. Charge Coupled Devices (CCDs) have a rich heritage in space-based imaging~\cite{GalileoCCD_1979,Trauger_HST_1990}, but CCDs with a skipper amplifier (skipper-CCDs) are a recent development that have never been deployed in the space environment. While conventional CCDs are limited to noise performance around $\sim$2\e, skipper-CCDs use repetitive non-destructive readout to measure the pixel charge down to sub-electron precision. This sub-electron noise performance has enabled low-threshold rare event searches with skipper-CCDs, and skipper-CCD experiments have set world-leading limits on sub-GeV DM-electron interactions~\cite{Sensei2023}. The Dark Matter Nano-satellite Equipped with Skipper Sensors (DarkNESS) mission will be the first deployment of skipper-CCDs in space, and will pave the way for future space-based implementations of skipper-CCDs.

This paper presents the development and status of the DarkNESS CubeSat Observatory designed to probe DM from Low Earth Orbit (LEO) using novel skipper-CCDs optimized to detect 1-10\,keV X-rays and provide sub-electron readout noise~\cite{Tiffenberg:skipper2017}. While in LEO, DarkNESS will investigate two DM signatures: X-rays produced by decaying DM and electron recoils from strongly-interacting sub-GeV DM. To probe decaying DM, DarkNESS will observe the diffuse X-ray emission from the Galactic Center to search for unidentified X-ray lines in the 1-10\,keV energy range.
Additionally, DarkNESS will leverage the sub-electron noise performance of skipper-CCDs to search for electron recoils from sub-GeV DM, exploring uncharted parameter space.

Laboratory validation of the DarkNESS subsystems is underway, with significant engineering work completed to establish the DarkNESS CubeSat design. Through designing an instrument payload that is compatible with a 6U CubeSat, DarkNESS will establish a platform for the future space-qualification of novel imaging sensor technologies.
With significant design reviews completed, DarkNESS is moving steadily toward a recently-awarded launch opportunity in late 2025. 
This paper will summarize the DarkNESS instrument design, highlight the technological milestones achieved during the design phase, and present the outlook for the DarkNESS mission's upcoming launch opportunity.
Sec.~\ref{sec:science} will present the mission's scientific goals. Sec.~\ref{sec:instrument} will describe the DarkNESS instrument, including its design specifications and considerations for operating skipper-CCDs in LEO. 
Sec.~\ref{sec:PDR} will describe the engineering components of the DarkNESS mission design, and Sec.~\ref{sec:plan} describes the upcoming integration, testing, and launch of the DarkNESS CubeSat Observatory.
Sec.~\ref{sec:conclusions} will present conclusions and the outlook of the DarkNESS mission.

\section{Scientific Objectives}
\label{sec:science}
Atmospheric attenuation can conceal certain signatures of DM, requiring space-based instruments to probe these DM models. DarkNESS will hunt for two such DM candidates from LEO, searching for electron recoils from strongly interacting sub-GeV DM and X-rays from DM decay.

\subsection{Strongly-interacting sub-GeV Dark Matter}


DM direct detection experiments are typically deployed in underground laboratories to shield the detectors from cosmogenic radiation and enable sensitivity to extremely small interaction cross-sections. However, if DM interacts strongly with ordinary matter, it will scatter in the Earth’s atmosphere and crust, attenuating the flux of DM reaching terrestrial detectors. At large interaction cross sections, the DM flux would not reach detectors deployed underground or on the Earth's surface. This scenario was examined in detail for sub-GeV DM in Ref.~\cite{emken:JCAP2019}. This work showed that strongly-interacting sub-GeV DM coupling to the Standard Model through an ultralight dark photon mediator could be a subdominant component of the cosmological DM.

\begin{figure}[t]
    \centering
    \begin{subfigure}[t]{.5\textwidth}
    \centering
    \raisebox{0.0\height}{\includegraphics[width=\linewidth]{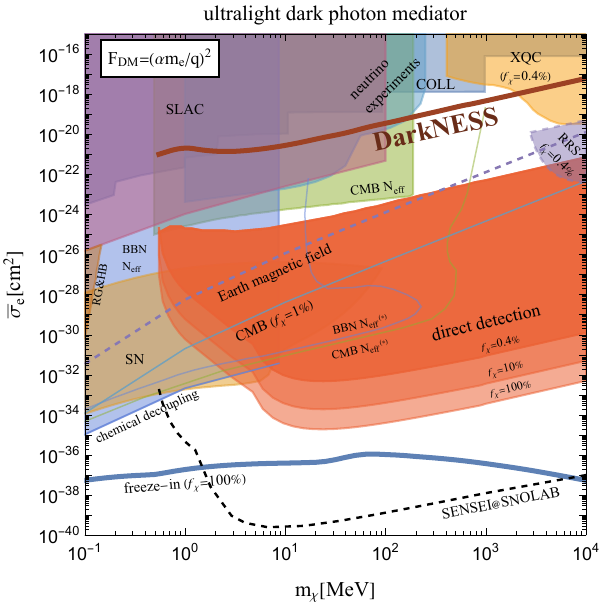}}
    \end{subfigure}%
    \begin{subfigure}[t]{.5\textwidth}
    \centering
    \raisebox{0.025\height}{\includegraphics[width = \linewidth]{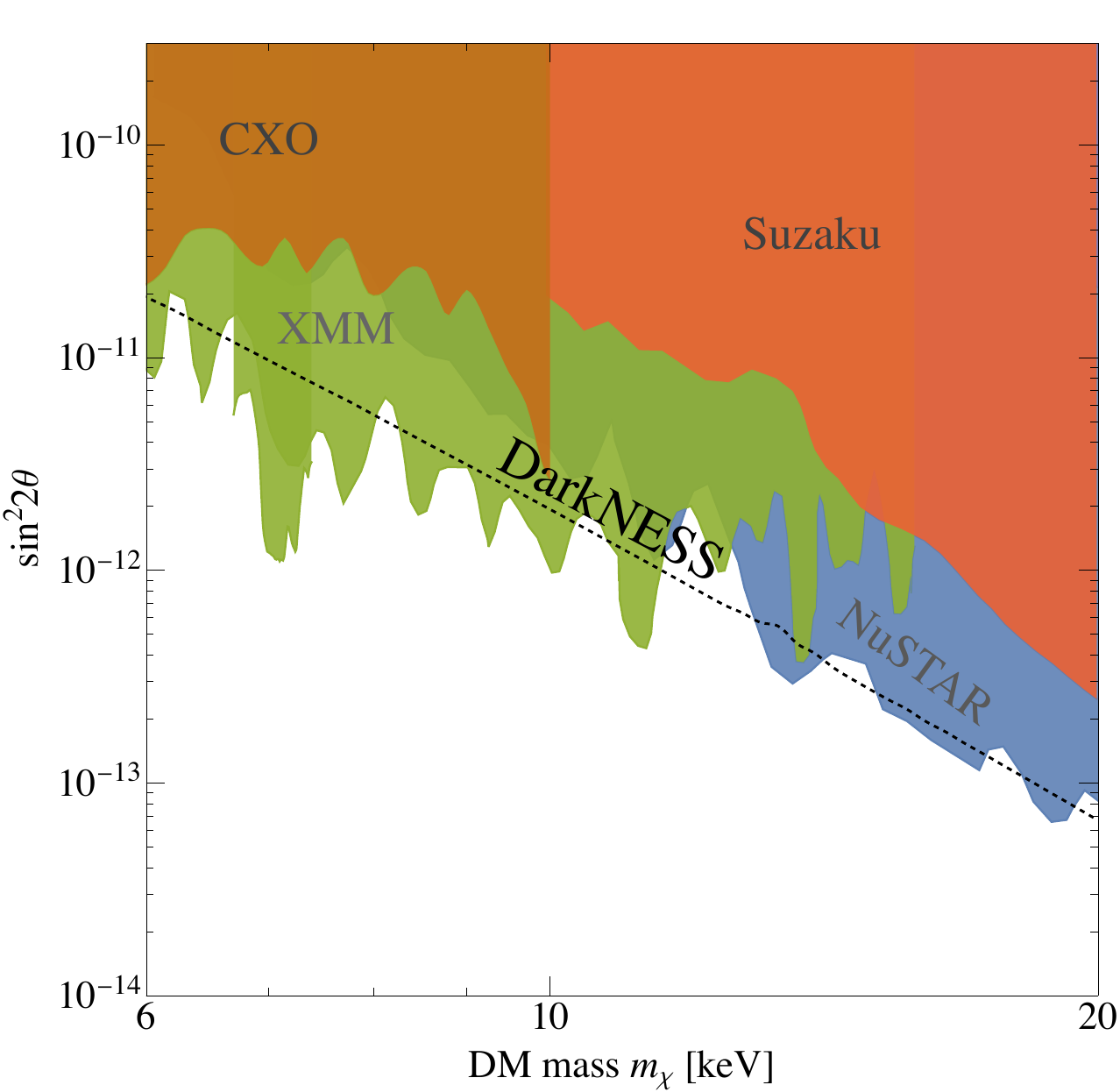}}
    \end{subfigure}%
\caption{\textit{Left:} Dark-matter electron interaction parameter space showing recent bounds on the DM-electron scattering cross section. DarkNESS will help expand the current direct detection exclusion bands (orange contour) up to higher cross-sections, constraining the parameter space for strongly-interacting sub-GeV DM. This DarkNESS discovery reach (red line) assumes that DM interacts via an ultralight mediator and a 0.1~gram-month exposure. See Ref.~\cite{Emken:2019tni} for more details. \textit{Right:} Exclusion limits on the DM decay rate into X-rays cast in units of $\sin^2 2\theta$ where $\theta$ is the mixing angle between the sterile and active neutrino. Black dashed line shows the projected DarkNESS 90\% C.L. upper limit using 25 hours of exposure time, assuming a Galactic Center background model from Ref.~\cite{figueroa}. The DarkNESS limit is compared with the best current limits from XMM (green)~\cite{XMM,Dessert_2020}, NuSTAR (blue)~\cite{NuStar}, Suzaku (red-orange)~\cite{Tamura_2015}, and CXO (orange)~\cite{Sicilian_2020}.\label{fig:Sat-sensitivity-1}}
\label{fig:DarkNESS-Science}
\end{figure}

DarkNESS will use skipper-CCDs with sub-electron noise to search for strongly interacting sub-GeV DM. To perform this search, DarkNESS will conduct low-noise observations pointing toward the constellation Cygnus. Cygnus lies near the Solar Apex, the direction in which the Sun moves as it orbits through the Galaxy's DM halo. This motion creates a `dark matter wind,' with DM particles appearing to flow predominantly from this direction relative to the CubeSat's perspective. Observing Cygnus aligns the instrument with this DM wind, maximizing the expected DM flux. As the CubeSat orbits the Earth, strongly interacting sub-GeV DM is expected to produce a large modulation signature in the instrument’s low-energy event rate due to the shadowing effect of the Earth~\cite{Kouvaris_2014}.

Using this technique, the discovery reach of DarkNESS to probe strongly interacting sub-GeV DM is shown as a red line in the left panel of Fig.~\ref{fig:Sat-sensitivity-1}. Using a 0.1 gram-month exposure, DarkNESS has the sensitivity to probe strongly-interacting sub-GeV DM in an unexplored region of the DM parameter space. The DarkNESS sensitivity will expand the current upper limits on the DM-electron scattering cross section to higher cross section values, filling the gap above terrestrial direct detection searches, represented by the orange shaded region in the left panel of Fig.~\ref{fig:Sat-sensitivity-1}. The reach of the DarkNESS upper limit is primarily constrained by the amount of shielding between the skipper-CCDs and the incoming DM flux, and the DarkNESS design uses minimal shielding ($\sim$100\,nm Al) to block stray light while maintaining sensitivity to strongly-interacting sub-GeV DM.
The DarkNESS instrumentation package includes an active mass of approximately 2 grams. Considering approximately 450 observations of Cygnus with a 10-minute exposure time and selecting 50\% of the exposed pixels after masking high-energy hits, DarkNESS can obtain a 0.1~gram-month exposure to achieve the discovery reach shown in Fig.~\ref{fig:Sat-sensitivity-1}.

\subsection{X-ray Signatures of Dark Matter}

DM decay is generically predicted in various particle DM scenarios, and photons are one of the promising channels to probe DM decay~\cite{Boddy_2022}. DM decay processes could give rise to a monoenergetic photon line, which would provide a smoking-gun signal of DM decay. Space-based observations are necessary to probe $\mathcal{O}\mathrm{(keV-GeV)}$ photons, since at these energies photons do not penetrate the Earth's atmosphere.

To probe decaying DM, the DarkNESS mission will search for unidentified X-ray lines in observations of the Galactic Center. During its operational lifetime, DarkNESS will have the opportunity to perform more than 1,200 X-ray observations of the Galactic center, each with a 15-minute integration time that can be split into shorter exposures to mitigate background radiation. This will result in a total exposure time of approximately 1\,Ms. The long exposure and wide Field of View (FOV) provide a large predicted signal rate from benchmark decaying DM models. For example, a sterile neutrino with $\mathcal{O}$(keV) mass decaying into an X-ray photon and an active neutrino arises from well-motivated extensions of the Standard Model and could make up the cosmological DM.

The decay of a ${\sim}$7\,keV sterile neutrino could explain the unidentified X-ray line at 3.5\,keV that was detected with high significance using stacked observations of galaxy clusters from instruments on the XMM-Newton satellite~\cite{Bulbul_2014}. This detection inspired a flurry of follow-up observations with mixed results (e.g.~\cite{Boyarsky:2014jta, Abazajian:2001vt, Abazajian:2014gza, Boyarsky:2018ktr}).
There is still some debate on how to interpret these conflicting results. The most recent results in this area disfavor the interpretation of the 3.5~keV line as resulting from DM decay (e.g.~\cite{Dessert_2020, Dessert:2023fen, Boyarsky_2020,2022swift, 2024TanDekkerDrlica}
) and suggest that the line was always an artifact rather than a valid signal. 
Using a 25\,h exposure, the expected DarkNESS sensitivity to sterile neutrino DM decaying to X-rays is shown in the right panel of Fig.~\ref{fig:Sat-sensitivity-1}.

\section{The DarkNESS Instrument}
\label{sec:instrument}
The DarkNESS mission integrates a skipper-CCD detector module into a 6U CubeSat to search for DM from LEO. 
Since the first demonstration of single-electron counting with a skipper-CCD~\cite{Tiffenberg:2017aac}, the SENSEI collaboration has developed skipper-CCDs for low-mass DM detection, exploiting the sensor's sub-electron noise and fine pixelation to set world-leading 
results in the field~\cite{Crisler:2018gci, SENSEI:2019ibb, SENSEI:2020dpa, senseibeam, Sensei2023}. Ongoing developments of the skipper-CCD technology further demonstrate the utility of sub-electron noise sensors for low-threshold rare event searches~\cite{2022oscura, OscuraSensors, OscuraEarlyScience}. 
The DarkNESS research program has focused on adapting skipper-CCDs and their readout electronics to the challenges of space-based operation so that they can be deployed in space. Through this effort, DarkNESS will open opportunities for the efficient testing and space certification of novel imaging sensors~\cite{Botti_2024,Sofo_Haro_2024,Lapi_2024}.

\begin{figure}[t]
    \centering
    \includegraphics[width=11cm]{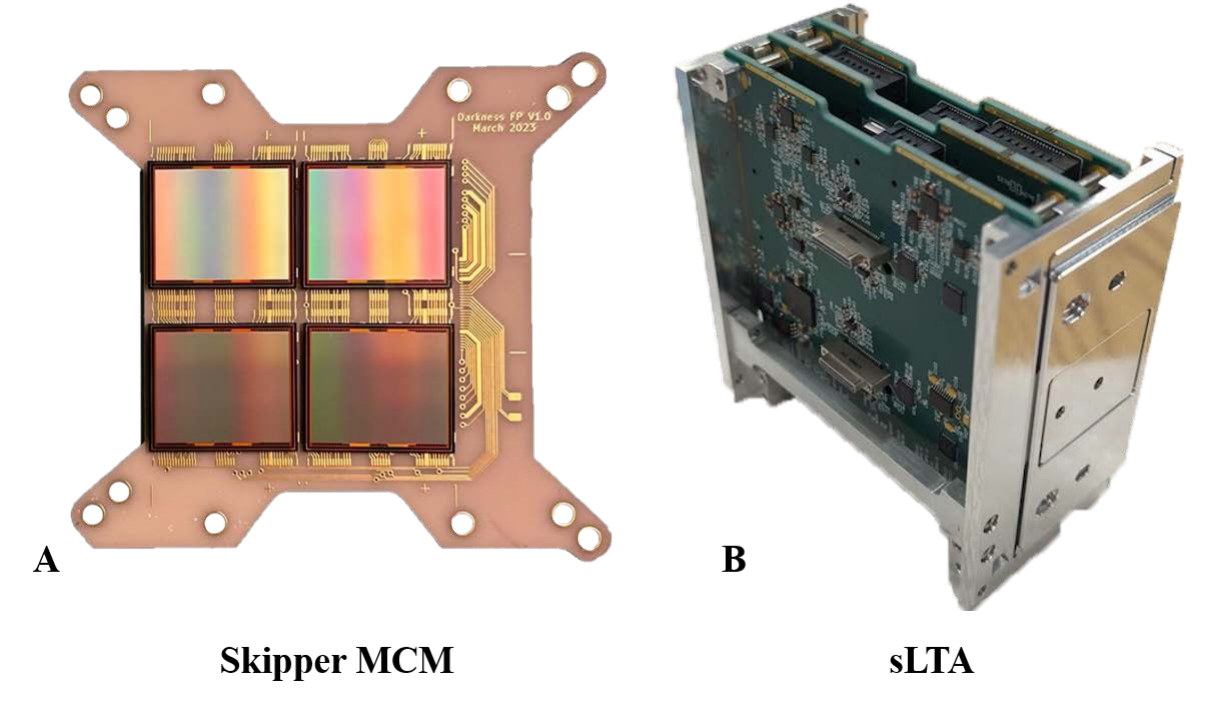}
    \caption{A) The Multi-Chip Module (MCM) designed, built, and tested for DarkNESS features the four 1.35 Mpix skipper-CCD designed by LBNL and fabricated at Microchip as part of the R\&D effort for Dark Matter experiments~\cite{OscuraSensors}. B) The space Low-Threshold Acquisition (sLTA) readout electronics housed in the DarkNESS thermal control board box~\cite{lta}.}
    \label{fig:instrument}
\end{figure}

\label{sec:payload}
\subsection{DarkNESS Multi Chip Module and Readout Electronics}
The DarkNESS instrument consists four 1.3\,Mpix skipper-CCDs integrated into a Multi Chip Module (MCM), as shown in Fig.~\ref{fig:instrument}~A). 
The skipper-CCDs were designed at the Microsystems Laboratory at Lawrence Berkeley National Laboratory (LBNL), and were recently fabricated as part of the Dark Matter research and development effort~\cite{OscuraSensors}. 
The LBNL design uses fully-depleted detectors up to 725~$\mu$m thick, enabling high efficiency of up to 10 keV~\cite{SteveCCD2003}. Specific aspects of the detectors are modified to apply the skipper-CCD technology to X-ray astronomy. A 500\,nm thick aluminum layer is applied to the front of the silicon detector to block visible and IR photons, serving as an X-ray entrance window. Backside processing to thin the detectors is currently being explored to optimize efficiency for X-ray applications.
Skipper-CCDs have demonstrated Fano-limited resolution for X-rays, as shown in Ref.~\cite{darioFano}. As a space-based X-ray spectrometer, the sensor packaging has also been designed for optimal performance. 

\begin{figure}[t]
    \centering
    \begin{subfigure}[t]{.56\textwidth}
    \centering
    \raisebox{0.05\height}{\includegraphics[width=\linewidth]{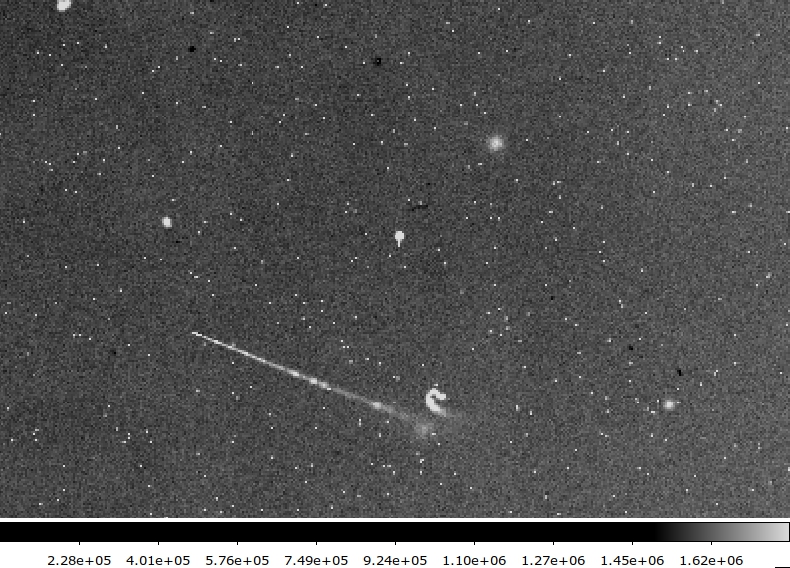}}
    \end{subfigure}%
    \hspace{0.03\textwidth}
    \begin{subfigure}[t]{.39\textwidth}
    \centering
    \raisebox{0.0\height}{\includegraphics[width = \linewidth]{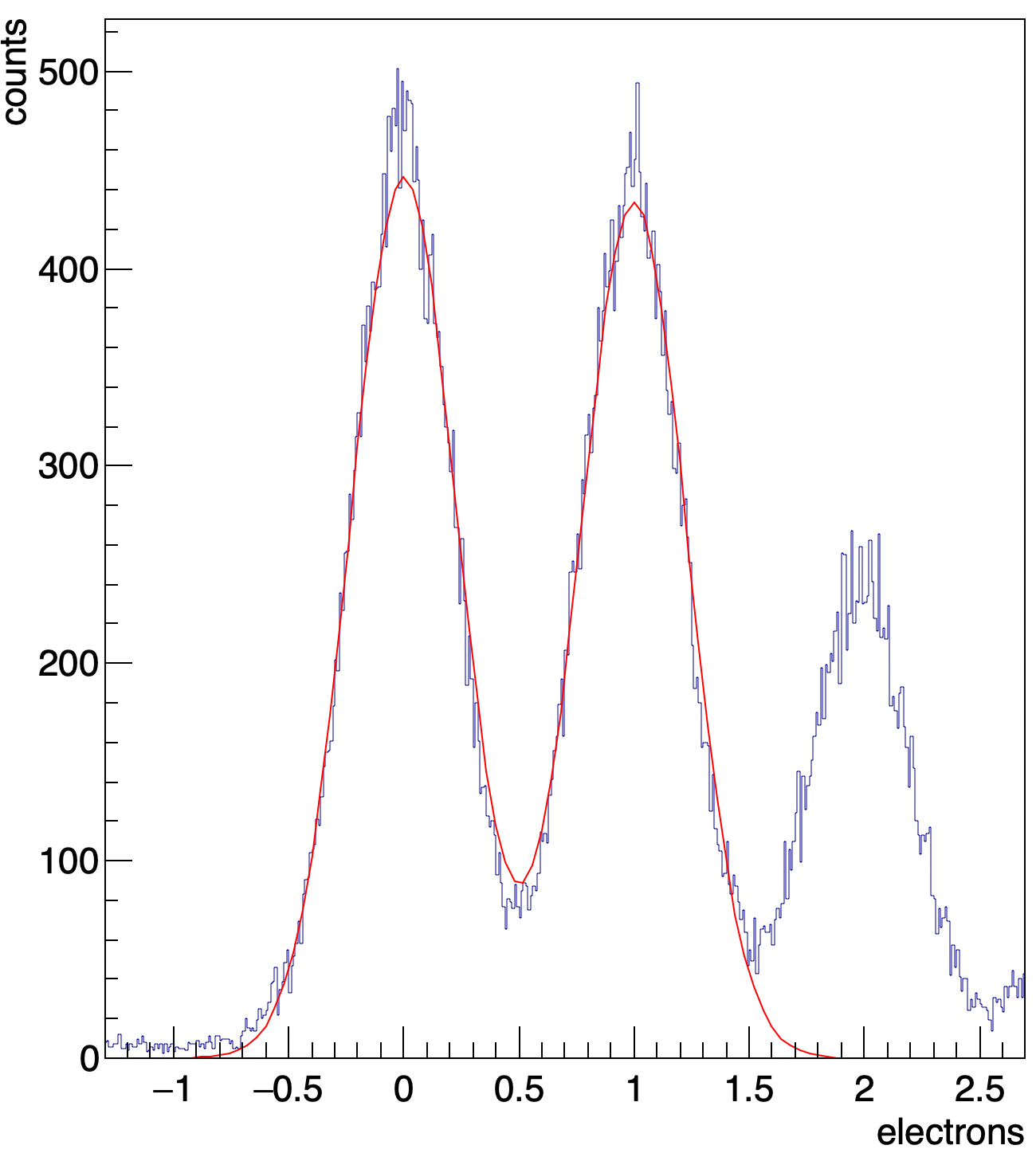}}
    \end{subfigure}%
\caption{\textit{Left:} An example of a test image collected at Fermilab with a DarkNESS skipper-CCD mounted on the MCM package is shown, and the readout with the sLTA is shown in Fig.~\ref{fig:instrument}, operating in a vacuum inside the chamber. The single-pixel hits show the ionization from 5.9 keV X-rays from a $^{55}$Fe source mounted inside the chamber. The longer track corresponds to a muon; other hits are likely multiple scattering electrons. This image was collected by measuring only a single sample for each pixel. \textit{Right:} Calibrated spectrum from prototype DarkNESS MCM that demonstrates of the photon counting capabilities using 300 skipper samples. The $x$-axis is in units of electrons; blue histogram shows the pixel spectrum, and the red line is a Poisson fit convolved with Gaussian noise. The individual electron peaks can be resolved with ~0.2\e of noise.}
    \label{fig:TestResults}
\end{figure}

The DarkNESS MCM uses a lightweight ceramic package to hold the four skipper-CCDs,
shown in Fig.~\ref{fig:instrument}. A custom flex circuit is epoxied and wire bonded to the MCM, connecting the MCM to the readout electronics. A prototype MCM and flex cable has been designed and tested in a cryogenic vacuum chamber at Fermi National Accelerator Laboratory (FNAL) using a $^{55}$Fe X-ray source. Fig.~\ref{fig:TestResults} shows an image and low-energy spectra from a prototype MCM.

\begin{figure}[th]
    \centering
    \includegraphics[height=5.8cm]{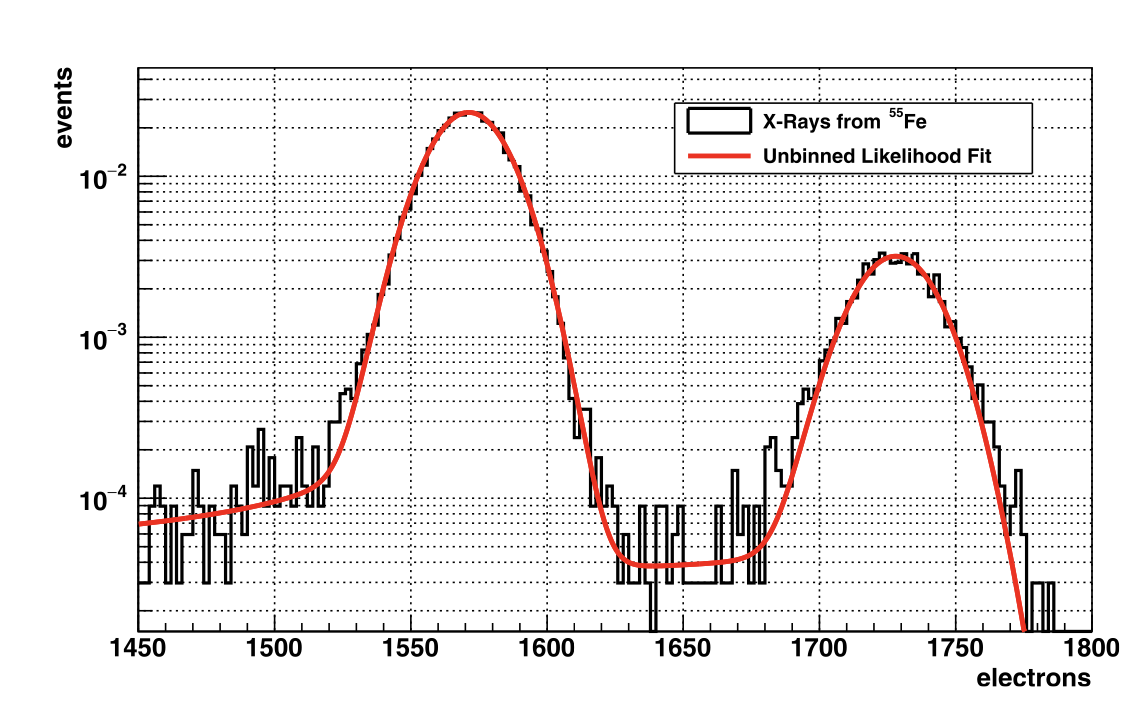}
    \caption{Energy spectrum of 5.9\,keV and 6.5\,keV X-rays from a $^{55}$Fe source measured with a skipper-CCD. The measured energy resolution is 50\,eV and is dominated by the Fano noise. Figure from Ref.~\cite{darioFanoFirst}.}
    \label{fig:newTech}
\end{figure}

The readout electronics for DarkNESS are based on the Low Threshold Acquisition (LTA) system developed for skipper-CCDs~\cite{lta} The LTA system provides the bias voltage to operate the skipper-CCD and controls the charge sequencing to read out the pixel array. To integrate the LTA into the 6U CubeSat platform, FNAL designed a compact version called the space-LTA (sLTA) shown in Fig.~\ref{fig:instrument}~B. The sLTA partitions the functionality of the LTA into three boards compatible with the CubeSat PC104 standard, uses less power than the standard LTA, and includes a copper plane for improved thermal management. The sLTA requires 10\,W of power to operate the skipper-CCDs and is controlled by a NanoAvionics payload controller through an ethernet connector. The sLTA is housed in an aluminum enclosure to aid in the electronics’ thermal management in the space environment.

The skipper-CCDs are cooled to 170\,K using a Ricor K508N compact rotary cryocooler~\cite{Mok2020_K508N}, and the CubeSat is designed to dissipate excess heat from the sLTA to the radiators. The instrument operating temperature requirement of 170\,K is necessary to the mitigate dark current due to thermal excitation of electron-hole pairs. Thermal management tests are ongoing to verify the sLTA thermal behavior in simulated space environment conditions, which are described in Sec.~\ref{sec:cryo}. DarkNESS X-ray observations require faster readout than direct DM applications, resulting in new requirements for the skipper-CCD front-end electronics. DarkNESS aims to achieve 
a readout speed of 250\,kPix/s, allowing full frame readout in 5\,s.


\subsection{DarkNESS Field of View and Expected Exposure}
\label{sec:FOV}
Given the CubeSat's volume constraints, the payload does not utilize X-ray optics. Each pixel will observe a 20-degree FOV defined by four circular apertures in front of the sensor array. As shown in the instrument design in Fig.~\ref{fig:PDRexterior} and Fig.~\ref{fig:PDRinstrument_CAD}, the instrument provides a total collecting area of 12\,cm$^2$. DarkNESS has a notably large field of view compared with other X-ray telescopes. For example, XMM EPIC-MOS has a FOV of $\sim$30~arcminutes and a collection area of 700\,cm$^2$ for 3\,keV X-rays. This means that one DarkNESS exposure observes a diffuse background flux comparable to 22 XMM EPIC-MOS images with the same exposure time. Since the skipper-CCDs can be fully depleted to a thickness of 725\,$\mu$m, increased efficiency for the 10--20\,keV X-rays is expected compared to thinner CCDs used in EPIC-MOS. The low-noise skipper-CCDs provide excellent energy resolution, as shown in Fig.~\ref{fig:newTech} ($\sigma \sim 50$\,eV at 6\,keV), limited by the Fano noise in silicon~\cite{Fano1947}.
The large FOV and Fano-limited energy resolution allow DarkNESS to achieve a high sensitivity to the diffuse X-ray background.

\subsection{Operating Skipper-CCDs in LEO}
\label{LEO}
The radiation environment in LEO is expected to be a significant challenge for detector operations. Laboratory work was performed with similar CCDs to understand their radiation tolerance~\cite{RadToleranceLBNL}, which indicated that the LBNL CCDs should perform well after exposure to space radiation. In order to ensure successful operation of skipper-CCDs in LEO, DarkNESS uses
Monte Carlo simulation tools and accelerated lifetime testing to assess the effects of operating skipper-CCDs in LEO.

\begin{figure}[t]
    \includegraphics[height=6cm]{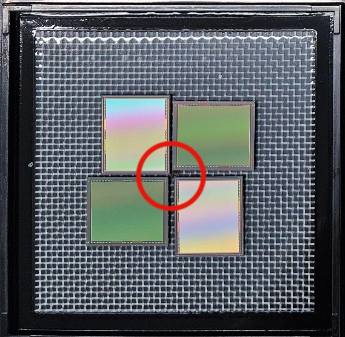}
    \includegraphics[height=6cm]{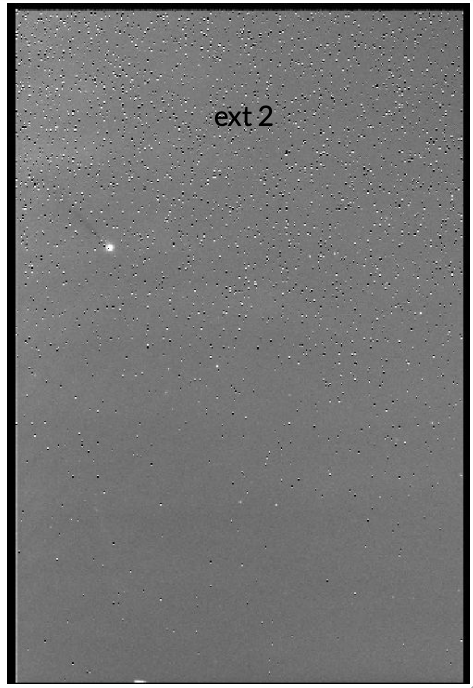}
    \centering
    \caption{Four CCDs in the sample holder are presented for the proton beam exposure (left). The proton beam was aligned with the center of the sample holder and covered approximately the area in the red circle. Results of single electron traps are seen for an image after pocket pumping~\cite{Janesick:2001}. The top of the image shows an area of the CCD exposed to the total dose, while the bottom of the image is further away from the proton beam (right).}
    \label{fig:protonsWarrenville}
\end{figure}

In order to evaluate the effect of radiation damage to the skipper-CCDs over the course of the mission, prototype DarkNESS skipper-CCDs were irradiated with a 217~MeV proton beam at the Northwestern Medicine Proton Center\footnote{\url{https://protoncenter.nm.org}}, delivering a total fluence of $1.2\times10^{10}$~protons/cm$^2$.
Despite this exposure, the skipper output stage showed no degradation, and the sensors maintained sub-electron noise performance~\cite{Roach_2024}. The robustness of the skipper amplifier enabled post-exposure studies of the density of single-electron traps in the imaging area, using the charge pumping technique described in~\cite{Janesick:2001}. These studies indicate that the protons created single-electron traps in the imaging area (see Fig.~\ref{fig:protonsWarrenville}). These defects were mainly divacancies, consistent with displacement damage observed in p-channel CCDs described in~\cite{Bebek:2002b, Hall:2017}. The radiation-induced trap density is $8\times10^4$/cm$^2$, corresponding to 0.18 traps per pixel, given the $15\times15~\mu$m$^2$ pixel size. This exposure was equivalent to a fluence of $3.4\times10^9$~protons/cm$^2$ for 12.5~MeV protons, roughly four times the total fluence expected over one year of DarkNESS operations. At a typical LEO altitude of 450~km, the expected trapped proton fluence for DarkNESS is $9\times10^8$~protons/cm$^2$ at 10~MeV\footnote{\url{https://www.spenvis.oma.be}}. Based on these results, we do not expect any degradation to the skipper amplifier over the course of the DarkNESS mission, but we note that approximately 5\% of the CCD pixels will accumulate single-electron traps per year in LEO. These traps can cause charge transfer inefficiency that 
requires masking techniques around high-energy events to eliminate spurious low-energy events during low-mass DM searches (such as the bleeding zone mask used in SENSEI DM analysis~\cite{SENSEI:2020dpa}).

Another challenge for operating skipper-CCDs in LEO is the generation of very low-energy hits (a few electrons) produced by ionizing radiation in the detector, which was discussed recently in Ref.~\cite{Gaido_2024}.
As highly-energetic charged particles traverse the detector, Cherenkov photons are produced that comprise a background for low-threshold direct DM searches~\cite{SecondaryRad_Rouven}. 
To mitigate this effect, DarkNESS will use an imaging analysis framework with tunable selection criteria that can define a region around high-energy events to be removed from the low-energy analysis (as done for the DM search in Ref.~\cite{SENSEI:2020dpa}).

\section{The DarkNESS Mission Design}
\label{sec:PDR}
\subsection{Concept of Operations and Mission-level Requirements}
\begin{figure}[t]
\includegraphics[width=0.90\linewidth]{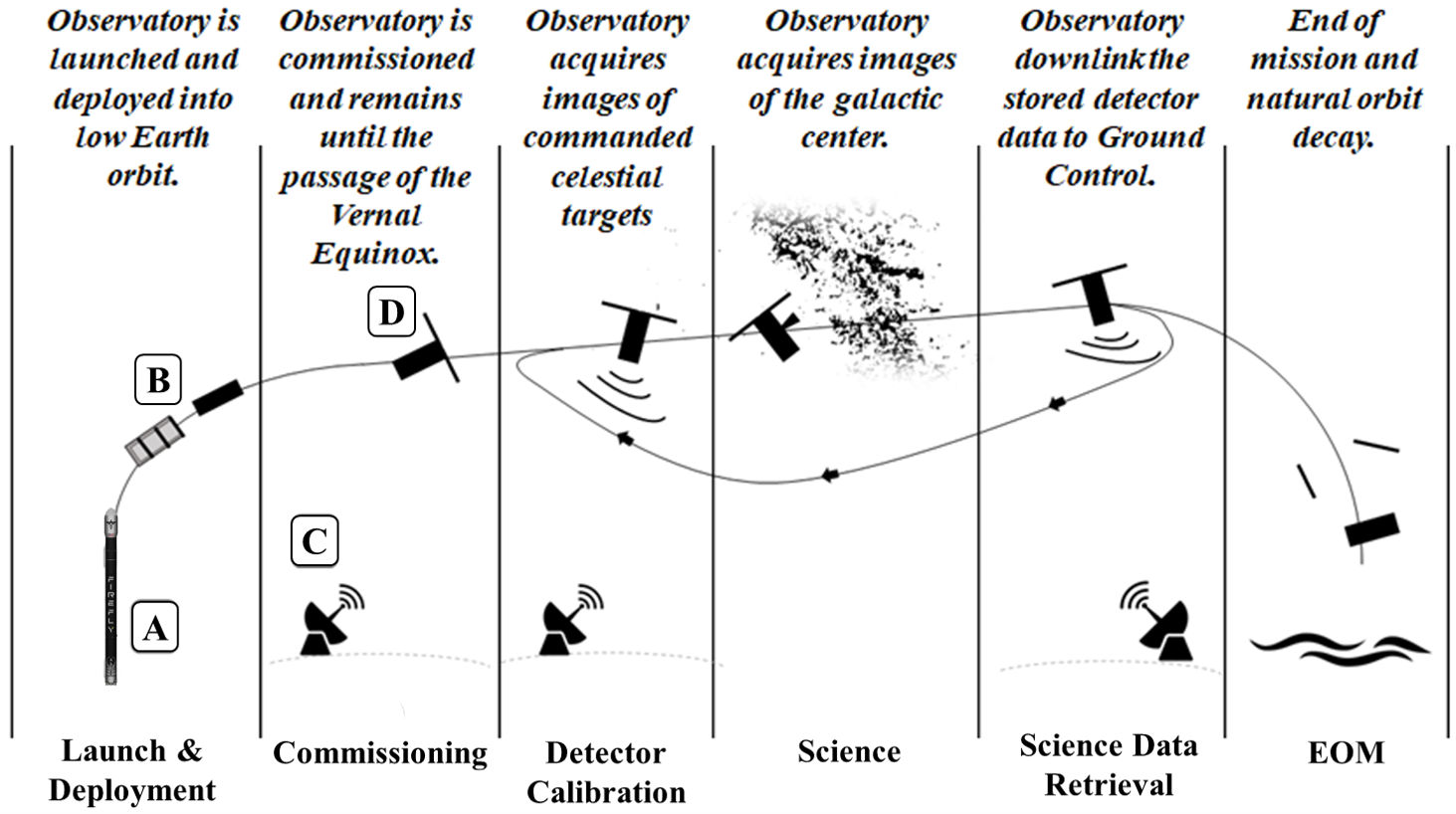}
\centering
\caption{DarkNESS concept of operations: The Launch Vehicle (A) delivers the Observatory to LEO, deployed by the  Dispenser (B). The Ground Station (C) configures the Observatory (D) subsystems. After commissioning, the Observatory begins instrument calibration and observations of the Galactic Center and Cygnus during science operations. Image data is transmitted to the Ground Station. The mission concludes with natural orbit decay and atmospheric re-entry within 5 years.}
\label{fig:conops}
\end{figure}

A requirements-driven systems engineering approach is applied to develop a CubeSat to achieve the science objectives described in Sec.~\ref{sec:science}. The Concept of Operations (ConOps) is illustrated in Fig.~\ref{fig:conops} and
defines system elements and the key mission functions of the mission architecture and guides the top-level mission requirements writing process. These requirements, stated in Table~\ref{tab:missiondrivers}, specify operational parameters and performance metrics to meet the science objectives.
The system functional requirements are categorized to create the trade space for subsystem sizing and constraints on the launch vehicle, dispenser, and ground station system elements. 
The CubeSat design process adopted a structured project life cycle review approach. Milestone reviews marked each phase of the project's formulation to evaluate technical progress, ensure alignment with mission requirements, and assess readiness for the next design phase. These reviews provided formal checkpoints to refine requirements, identify risks, and verify design maturity.

The ConOps shown in Fig.~\ref{fig:conops} describes how the mission meets its objectives using the system architecture. After launch and deployment in LEO, the CubeSat performs initial subsystem commissioning before entering the science phase. Following the Vernal Equinox, it autonomously points at celestial targets for 15-minute exposures. It also manages instrument thermal conditions and downlinks science data and telemetry to the ground station for analysis and archiving. The mission will end with the natural orbit decay of CubeSat.

\begin{table}[t]
    \centering
    \caption{DarkNESS Mission-level Requirements}
    \begin{tabular}{ |p{0.5cm}||p{12cm}|  }
    \hline
    \multicolumn{1}{|c||}{\textit{Category}} & \multicolumn{1}{|c|}{\textit{Requirement Statement}} \\
    \hline
    \hline
    \multicolumn{1}{|c||}{\textit{Science}} & \textit{\textbf{MIR-1:}} DarkNESS minimum mission life shall be one year. 
    
    \textit{\textbf{MIR-2:}} The primary science objective shall be achievable within the period marked by the Vernal Equinox and terminating on the Autumnal Equinox. \\
    \hline
    
    \multicolumn{1}{|c||}{\textit{Orbit}} & \textit{\textbf{MIR-3:}} DarkNESS orbit shall be bound between mid-inclination and Sun-synchronous.
    
    \textit{\textbf{MIR-4:}} DarkNESS orbit shall be in the range of 400-550\,km in altitude. \\
    \hline
    \multicolumn{1}{|c||}{\textit{Instrument}} & \textit{\textbf{MIR-5:}} The instrument shall be powered at 10\,W during umbral passage.
    
    \textit{\textbf{MIR-6:}} The cryocooler shall operate at 11\,W with a duty cycle of 100\%. \\
    \hline
    \multicolumn{1}{|c||}{\textit{Data}} & \textit{\textbf{MIR-7:}} The maximum data size for each raw image produced by the instrument shall be 700\,MB.
    \newline\textit{\textbf{note:}} \textit{the expectation is one downlinked image per }\textit{\textbf{DAY}}\textit{ during the science season (Vernal Equinox to Autumnal Equinox, MIR-2).}
    
    \textit{\textbf{MIR-8:}} The maximum data size for each histogram dataset produced by the instrument shall be $\sim$2.5\,kB.
    \newline\textit{\textbf{note:}} \textit{the expectation is one downlinked histogram per }\textit{\textbf{ORBIT}}\textit{ during the science season (Vernal Equinox to Autumnal Equinox, MIR-2).} \\
    \hline
    \multicolumn{1}{|c||}{\textit{ADCS}} &  \textit{\textbf{MIR-9:}} Course attitude knowledge and 3-axis control. 
    
    Pointing Modes: 
        \begin{enumerate}[nolistsep,nosep]
            \item \textbf{Sunlit Orbit:} Solar panel to sun tracking (primary) / Radiator to dark space (secondary)
            \item \textbf{Umbral Science:} X-ray detector to inertial target (primary) / Radiator to dark space (secondary)
            \item \textbf{Ground Station Tracking:} S-band radio to GS (primary) / Instrument (+Z) to ram (secondary)
        \end{enumerate} 
        \\
    \hline
    \multicolumn{1}{|c||}{\textit{Thermal}} & \textit{\textbf{MIR-10: }}Passive thermal system components shall include two frame-mounted radiator panels and thermal straps transporting heat from the active cryocooler. 
    
    \textit{\textbf{note: }}Ricor K508N cryocooler removes heat from the focal plane assembly for transport via passive elements to the frame-mounted radiator panels. 
    \\
    \hline
    \end{tabular}
    \label{tab:missiondrivers}
\end{table}

\subsubsection{Observatory Configuration}
\begin{figure}[t]
    \centering
    \includegraphics[width=0.95\linewidth]{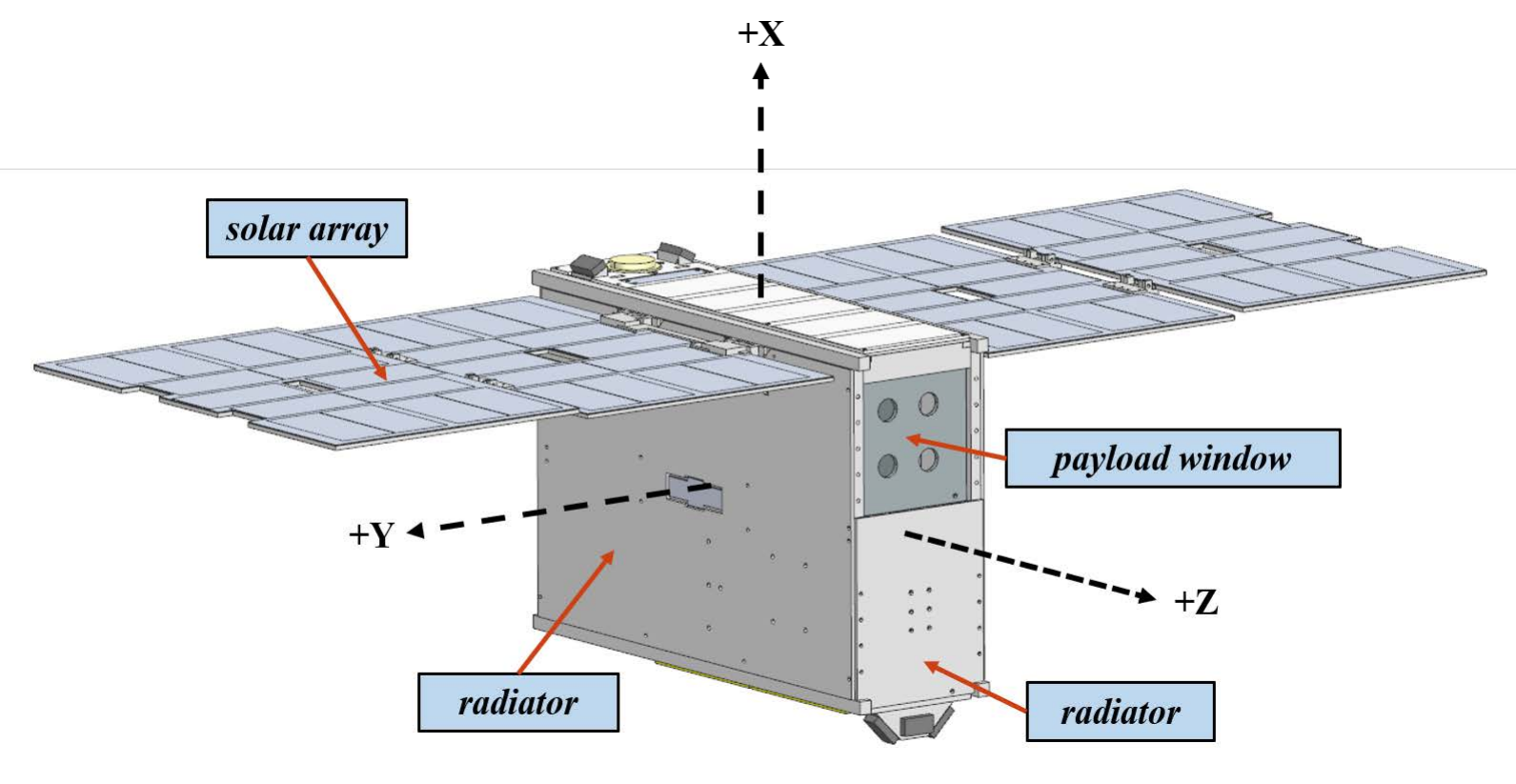}
    \caption{Key CubeSat features include three body-mounted radiator panels, a dual-deployable solar array, and the instrument aperture configured on the NanoAvionics 6U platform.}
    \label{fig:PDRexterior}
\end{figure}

With basic requirements set in hand, Kongsberg NanoAvionics (KNA) was selected to provide a flight-proven 6U CubeSat platform to satisfy the mission requirements. A key feature of the configuration proposed by KNA is the large, dual-deployable solar arrays. As stated in Table~\ref{tab:missiondrivers}, the payload consumes $\sim$20\,W when operating at 15-minute durations once per umbral passage. When summed with the other system power requirements, $\sim$30\,W of daily average orbit power must be provided to satisfy mission needs. 
Another key CubeSat feature is the thermal rejection system of three body-mounted radiator panels that provide passive thermal management for the instrument, which consists of a compact cryocooler (see Sec.~\ref{sec:cryo}) that chills the skipper-CCDs down to 170\,K. The detector assembly has four apertures on the forward-facing panel of the CubeSat and is designed for FOV requirements discussed in Sec.~\ref{sec:FOV}. Fig.~\ref{fig:PDRexterior} illustrates these key external CubeSat features.

\begin{figure}[t]
    \includegraphics[width=0.8\linewidth]{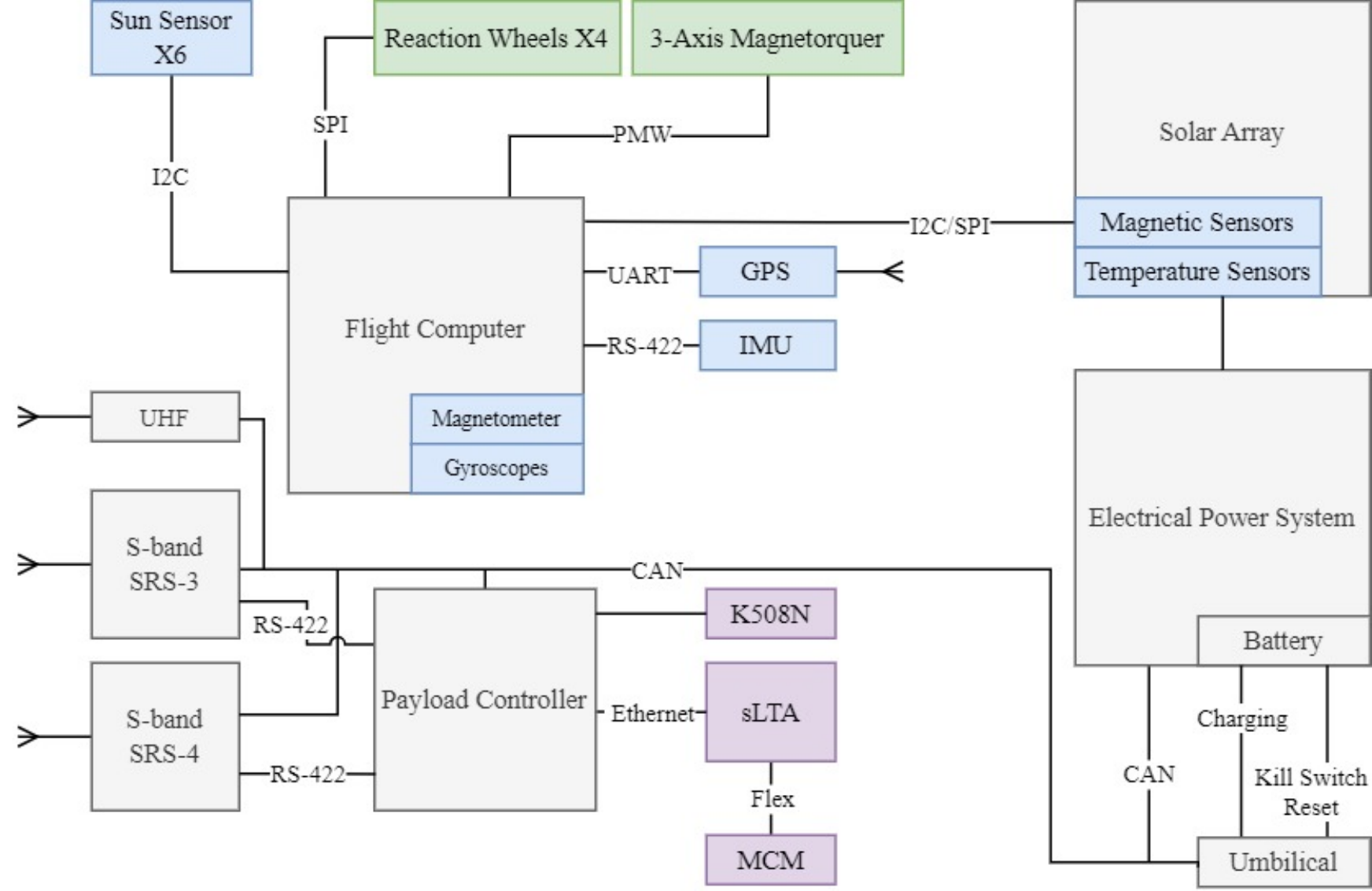}
    \centering
    \caption{DarkNESS subsystem configuration proposed by KNA. Subsystems are shown with the key data interfaces to the payload. The control actuators (green) provide pointing of the CubeSat to modes described in Table~\ref{tab:missiondrivers}. Sensor components (blue) used in attitude determination include inertial measurement unit, sun sensors, magnetometers, gyroscopes, and GPS for position tracking}
    \label{fig:subsystems}
\end{figure}

\begin{figure}[t]
    \includegraphics[width=0.9\linewidth]{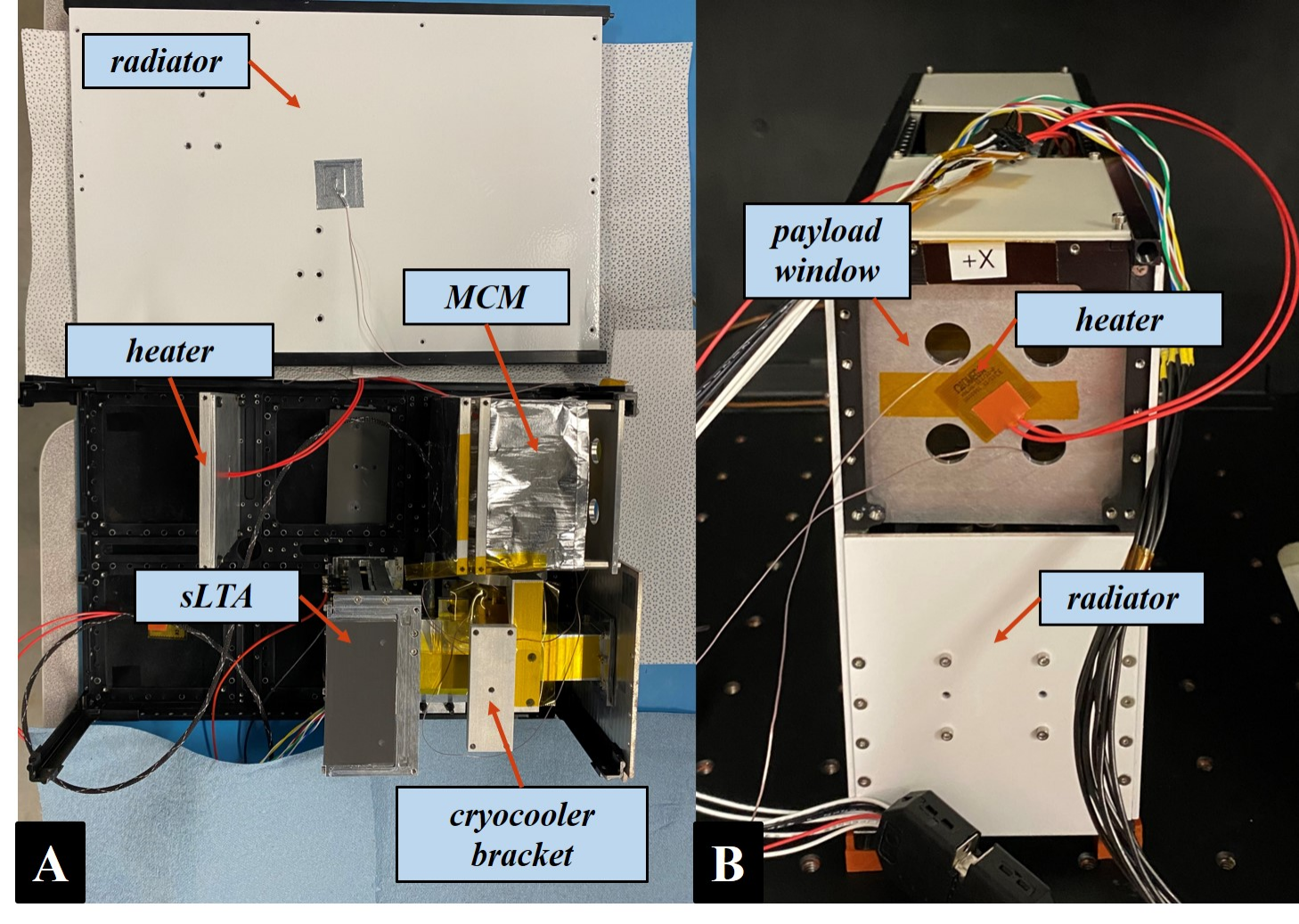}
    \centering
    \caption{Left: Design of the DarkNESS CubeSat configuration. The 6U CubeSat has dimensions 30\,cm x 20\,cm x 10\,cm. Right: Engineering model used at LASSI-UIUC to test the thermal design for DarkNESS. The model has the cryocooler, readout electronics, payload detector assembly (with dummy sensors), and radiator panels with surface treated for thermal emissivity (from Aeroglaze).\label{fig:thermaltest}}
    
\end{figure}

\begin{figure}[t]
    \centering
    \includegraphics[width=0.9\linewidth]{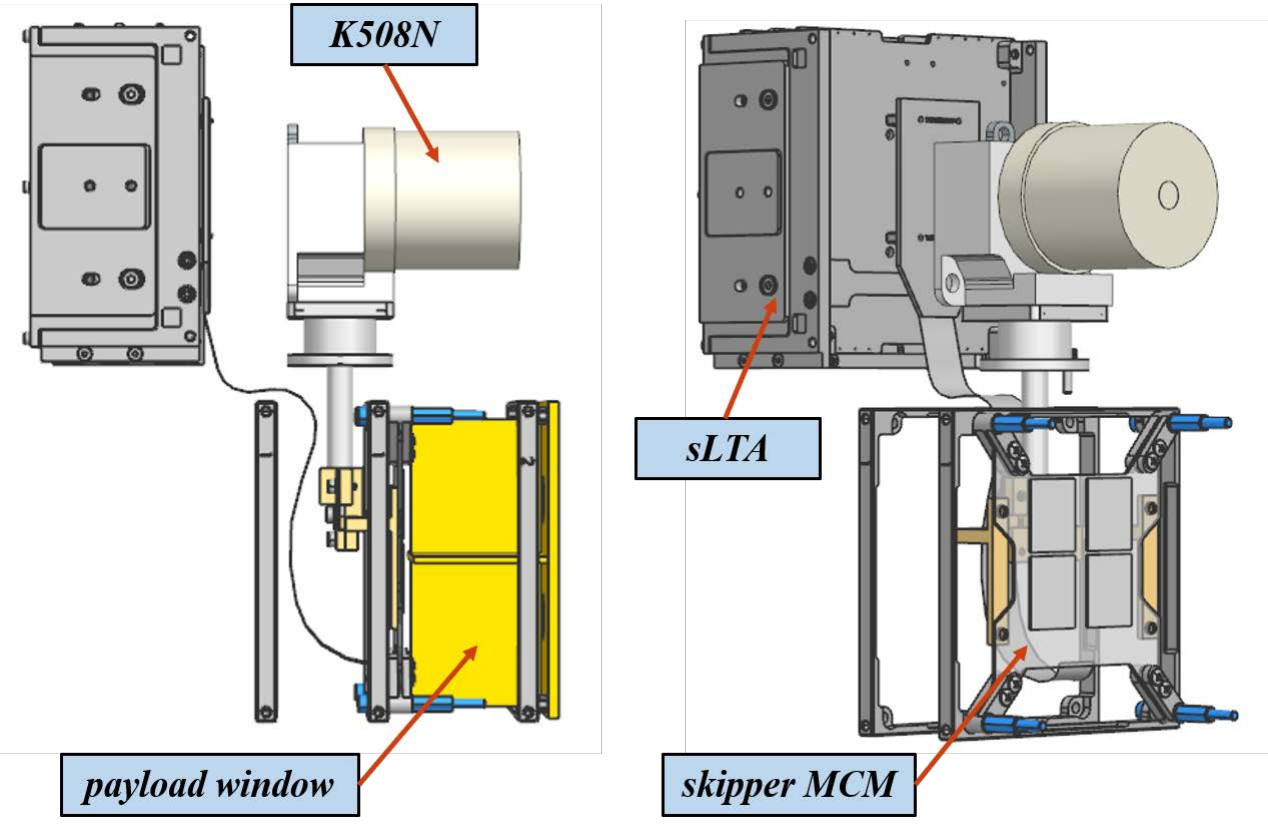}
    \caption{The instrument assembly with the Ricor cryocooler interfaced to the Multi-Camera Module (MCM) composed of four skipper-CCDs. sLTA electronics are thermally managed by an aluminum board box that interfaces two radiator panels.}
    \label{fig:PDRinstrument_CAD}
\end{figure}

The payload controller integrates software functions and interfaces for instrument data storage and transmission using dedicated S-band radios, as depicted in Fig.~\ref{fig:subsystems}. Three radios are provided: one UHF radio manufactured by KNA and two Satlab S-band radios (SRS-3 and SRS-4). The UHF radio, operating at 301-304 MHz, receives ground station commands to operate the CubeSat. The high-power SRS-4 radio handles raw image data and CubeSat telemetry downlinks at 2200-2290 MHz, while the low-power SRS-3 serves as a backup at a reduced data throughput. All radios maintain positive link margins for pass elevations between 5 and 90 degrees, with a 3 dB margin. The link budget was calculated for an assumed altitude of 510 km.

The CubeSat’s orientation is controlled using attitude sensors such as magnetometers, sun sensors, gyroscopes, and an Inertial Measurement Unit (IMU). Sensor data is integrated via a Kalman filter and combined with ADCS software developed by KNA, providing 3-deg attitude accuracy. Attitude control is managed with reaction wheels, while stored momentum is dumped using magnetometers to achieve 5-degree pointing accuracy. The detector’s large FOV drives coarse pointing requirements. The ADCS software operates on the KNA-designed flight computer, supporting multiple pointing modes as defined in Table~\ref{tab:missiondrivers}. A GPS receiver provides real-time orbital parameters and includes a Pulse Per Second (PPS) signal for precise time synchronization.

\subsubsection{Payload Thermal Testing}
\label{sec:cryo}



Thermal management posed design challenges for integration on a 6U CubeSat. The skipper-CCDs were attached to the MCM ceramic substrate (see Fig.~\ref{fig:instrument}). Continuous active thermal management is required to maintain operational readiness during umbral passages.

As observed during laboratory testing, the large ceramic surface area requires lengthy cool-down periods to reach 170\,K. Radiative heat loads, both internal and external, reduce the cryocooler’s cooling power. Additionally, heat from the K508N’s primary thermal interfaces must be dissipated. Four main thermal dissipation surfaces are offloaded to the external radiator panels to condition the cryocooler and detailed in Ref.~\cite{Alpine2023}.

Aluminized mylar shielding reduces the radiative heat load from the CubeSat interior on the detector. This reduces internal radiation reaching the MCM ceramic substrate as a thermal load on the cryocooler’s cold tip. A custom copper bracket passively transfers heat from the cryocooler to body-mounted radiator panels. At the same time, the sLTA readout electronics (approximately 10\,W) are similarly managed through passive transfer to these radiators. The radiators are coated with Socomore A276 white paint to limit spectral absorption from solar radiation. Attitude control constraints are imposed to prevent direct exposure of the instrument window to solar and albedo radiation. Early thermal tests, including a full-scale engineering thermal model, have demonstrated the feasibility of maintaining the detector at 170 ± 10\,K (see Ref.~\cite{Alpine2023}). Ongoing testing has refined the cryocooler configuration and passive thermal component designs.

\begin{figure}[t]
    \includegraphics[width=0.9\linewidth]{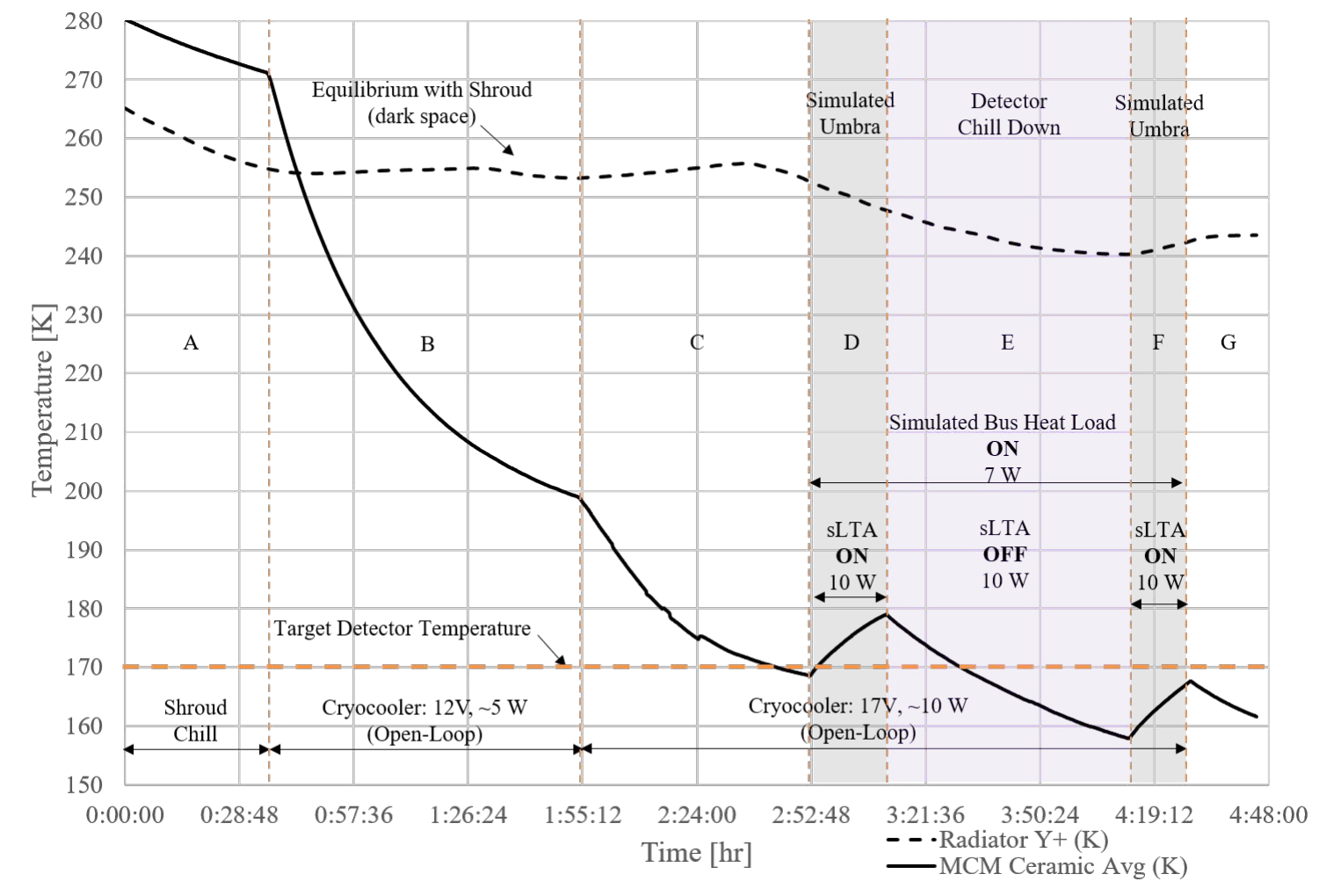}
    \centering
    \caption{Results from TVAC tests at LASSI-UIUC. The chamber was maintained at $5.0\times 10^{-4}$\,Torr and the shroud at 255\,K (Stage A). The cryocooler (10 W) gets the MCM to 168\,K (Stage C). The bus heat load is added (7\,W), and the cycles of 15-minute sLTA readout with the additional load (10\,W) are shown in Stages D and F. The MCM stays around the target operating point 170\,K. These tests were done with open-loop control for the cryocooler. Closed-loop cryocooler control will be implemented to reduce the temperature fluctuations around the operating point.\label{fig:thermalResults2024}}
\end{figure}

Hardware testing of the thermal control system was undertaken as part of early risk-mitigation efforts. A flight structural frame was acquired from KNA for instrument integration and testing under thermal vacuum (TVAC) conditions at the University of Illinois. A mechanical MCM featuring four skipper-CCD packages mounted to a ceramic substrate (Fig.~\ref{fig:instrument} and Fig.~\ref{fig:PDRinstrument_CAD}) was used alongside a functional sLTA prototype housed in an aluminum board box for thermal management. The instrument was assembled into the structural frame with passive thermal hardware. The cryocooler was mounted using a custom thermal bracket design, with its cold tip interfacing the ceramic substrate via a custom copper interface, as shown in Fig.~\ref{fig:PDRinstrument_CAD}.

The instrument and thermal hardware components are configured into the 6U structural frame in Fig.~\ref{fig:thermaltest}. Thermocouples were attached to key surfaces of the cryocooler, MCM ceramic, and radiator panels to observe the thermal response. Two Kapton heaters were also integrated into the experiment to simulate external heating of the payload window (2\,W) and the internal platform components (10\,W). Thermal testing was performed inside a thermal vacuum chamber equipped with a liquid nitrogen shroud, which was brought to a high vacuum using a roughing and turbo pump setup.

The thermal response of the instrument detector is shown in Fig.~\ref{fig:thermalResults2024} over a 5-hour test. The average MCM ceramic temperature of three thermocouples placed on the front surface is divided into phases representing vacuum and power conditions. The chamber was cooled to $\sim$255\,K using a liquid nitrogen shroud before powering the cryocooler at 5\,W in Phase B. The chamber was manually controlled and kept warmer than the expected space environment. In Phase C, the cryocooler was set to 10\,W, bringing the detector to the target 170\,K. Powering the sLTA at 10\,W in Phase D introduced internal radiation to the MCM ceramic, causing a temperature increase. The sLTA was off in Phase E, while a 7\,W CubeSat avionics heater was activated. The cryocooler was allowed to cool the ceramic below 170\,K before re-powering the sLTA in Phase F. A 15-minute duration passed without the ceramic temperature rising above 170\,K, simulating an umbral passage observation. The shroud was cooled further to 240\,K to simulate umbra conditions but remained warmer than expected. The test concluded in Phase G, when the liquid nitrogen was depleted.

\subsection{Orbit Considerations}
\label{sec:orbit}

The primary science objective targeting the Galactic center is constrained by celestial geometry, driven by seasonal and orbital factors. For half of Earth's yearly orbit, the Sun’s position and brightness can interfere with observations from the instrument’s wide FOV. 

\begin{figure}[t]
    \centering
    \includegraphics[width=0.9\linewidth]{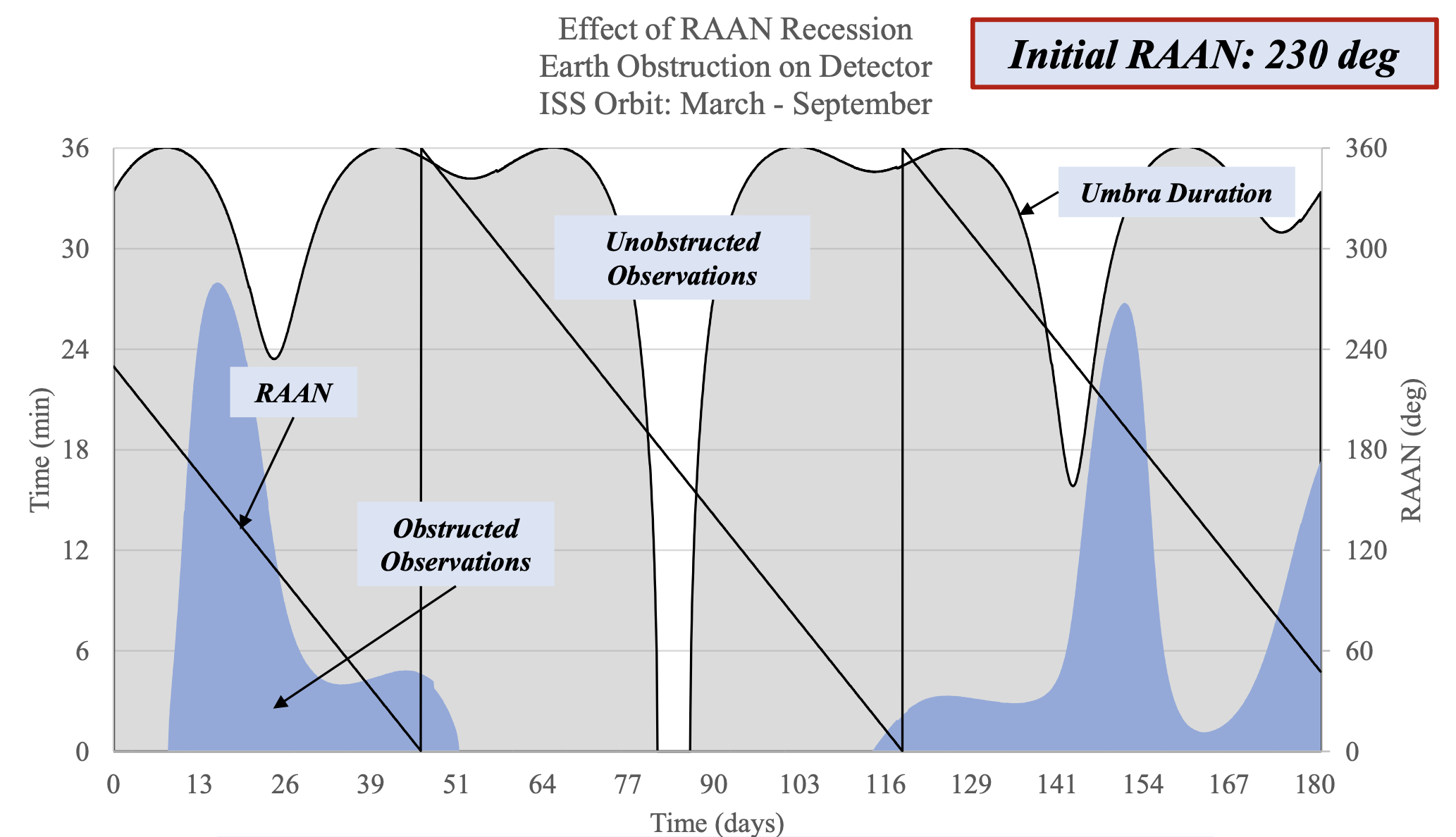}
    \caption{Earth obstruction considerations for imaging the galactic center with DarkNESS with an initial RAAN of 230 degrees. 
    }
    \label{fig:orbit}
\end{figure}

Prior to selecting a launch vehicle, the orbital inclination of the CubeSat was not known. Accordingly, investigations were undertaken to determine the impact of orbital inclination on mission operations. For mid-inclination orbits, the Right Ascension of the Ascending Node (RAAN) affects the alignment between the instrument’s FOV and the Earth, occasionally causing the detector to be obscured by the Earth’s limb while observing the Galactic Center and Cygnus. This solar interference does not happen between the Vernal and Autumnal Equinoxes, allowing scientific observations to be made.

Mission analysis quantified Earth’s obscuration of the instrument FOV during umbral passage. This data aids operations planning and informs precise timing commands for initiating the instrument pointing toward the Galactic center. After determining observation windows, careful scheduling reduces albedo exposure, lowering the heat load applied to the skipper-CCD array from Earth’s reflected solar and infrared radiation during obscuration events. 
The duration of the umbra is directly related to the RAAN precession of the orbit and is dominated by the second zonal harmonic of Earth’s gravitational potential, particularly for mid-inclination orbits. Because the day-of-flight RAAN is not yet known, the analysis is repeated for a range of possible initial RAAN conditions. An example of obscuration data with a day-of-flight RAAN of 230 degrees is presented in Fig.~\ref{fig:orbit}. A scheduling tool for DarkNESS operations is currently under development and features obscuration computations to provide schedules for DarkNESS operations.


\section{DarkNESS Integration, Testing, and Launch}

 \label{sec:plan}
 DarkNESS is entering the development lifecycle's system assembly, integration, and test phase. Verification of the design to meet industry requirements for space deployment, particularly with the launch vehicle and dispenser (see Sec~\ref{sec:launch}), is also underway.

\subsection{Design Completion Review}
\label{sec:CDR}
The DarkNESS program is progressing through the final design phase, which sets the baseline for the program's implementation stage. Tests using an engineering model of the instrument also continue in thermal vacuum conditions, while structural simulations verify that the design meets Firefly Alpha launch load requirements (see Sec.~\ref{sec:launch}). Key upcoming deliverables include subsystem test plans, updated design documentation, and results from trade studies, ensuring compliance with the verification control requirements. Mission operations plans, including commissioning and science data collection, are being formulated in detail.

\subsection{Manufacturing and Assembly}
\label{sec:build}

The DarkNESS manufacturing and assembly phase centers on integrating the skipper-CCD instrument with KNA's commercial 6U CubeSat platform subsystems. Configurations and modifications are made to the CubeSat’s power distribution subsystem and payload controller to meet the specific interfacing requirements with the instrument defined during the design phases. Before assembly, each subsystem is tested based on procedures described in the subsystem test plans to verify functionality.

Manufacturing plans for KNA-produced subsystems, including the electrical power system, flight computer, and structural bus frame, follow an established schedule that aligns with the mission’s critical milestones. Long-lead items, such as the cryocooler and skipper-CCD electronics, are procured early to reduce risk and prevent delays during integration. 
Close coordination between the manufacturing teams and suppliers helps ensure all components are delivered on time, minimizing risks and maintaining the mission’s integration schedule.

\subsection{Dispenser and Launch}
\label{sec:launch}
Firefly Aerospace awarded DarkNESS a launch opportunity through its DREAM 2.0 program\footnote{https://fireflyspace.com/dream/}, which supports educational and scientific CubeSat missions by providing free rideshare access to space on Firefly’s Alpha launch vehicle. As a rideshare payload, DarkNESS must be a ``Do-No-Harm'' satellite, meaning it cannot impose any mission architecture requirements on the host mission. The launch is expected no earlier than 4Q2025, though specific orbit parameters have yet to be determined. The DarkNESS mission was selected for its scientific contributions, aligning with DREAM’s goals of fostering innovation in space research. Coordination with Firefly is ongoing to ensure compatibility with their launch systems and requirements.
Mission analysis and subsystem trade studies have been intentionally scoped to accommodate uncertainty in mission orbit parameters for any launch opportunity. As a result, DarkNESS can accommodate a wide range of potential orbit parameters that will be finalized as the launch manifests with the primary host mission. Imposing wide-range orbit bounds on the mission analysis and the subsystem trade spaces proved fruitful in establishing an orbit-agnostic design. 
CubeSat payloads in rideshare launches require dispensers to eject them from the launch vehicle. While many commercial dispensers exist, not all accommodate CubeSat-specific mechanical constraints. DarkNESS, with dual-deployable solar arrays overlaying thick radiator panels, requires a dispenser with sufficient protrusion clearance. Exolaunch’s NOVA 6U/8U CubeSat dispenser meets these needs, allowing up to 25 mm protrusions, compared to the CubeSat Design Specification limit of 6.5 mm. This expanded volume is critical to successfully deploying DarkNESS in orbit.

\subsection{Deployment and Mission Operations}
\label{sec:operations}
DarkNESS will follow a sequence of pre-defined events after deployment from the Exolaunch Dispenser. Electrical inhibits are released upon ejection, enabling the electrical power system to turn on. The flight computer is started and executes its post-deployment tasks, including transmitting telemetry via the omni-directional UHF radio. This telemetry provides status on subsystems, position, orientation, and sensor measurements. Additionally, the flight computer activates the Attitude Determination and Control System (ADCS) in detumbling mode, using magnetorquers to stabilize the CubeSat and eliminate any rotation from deployment. Mission operations start when the CubeSat first flies over the Ground Station. During this critical phase, the ground station receives the intermittent beacon signal identifying the CubeSat and then establishes communication with it to configure its operational state, marking the beginning of the commissioning phase. Once completed, DarkNESS will transition into the science phase, performing observations of the Galactic Center and Cygnus and downlinking image data for analysis and archiving. DarkNESS expects at least one year of science operations before decommissioning due to natural orbit decay.

\section{Conclusions and Outlook}
\label{sec:conclusions}
The DarkNESS mission aims to deploy skipper-CCDs on a 6U CubeSat to probe unexplored dark matter parameter spaces from LEO. This novel approach leverages cutting-edge advancements in sensor technology and compact readout electronics supported by commercial cryocoolers. The mission presents a unique opportunity to perform pioneering science in space, combining recent innovations in skipper-CCD technology with compact and efficient sLTA electronics to achieve sub-electron noise for X-ray astronomy and DM searches. The mission has made significant progress in subsystem development, including sensors, readout, and thermal control.
As DarkNESS advances toward its launch, it stands to pave the way for future space-based instruments utilizing skipper-CCDs for X-ray spectrometry and single-photon counting applications.

\section{Acknowledgement}
\label{sec:acknowledgement}
We are grateful for the support of the Heising-Simons Foundation. This work was supported through the Fermilab LDRD program. PA acknowledges support from the DOE SCGSR program. This work was supported by Fermilab under U. S. Department of Energy (DOE) Contract No. DE-AC02-07CH11359.

\clearpage
\newpage
\addtocontents{toc}{\protect\setcounter{tocdepth}{1}}

\section*{}\label{sec:references}\addcontentsline{toc}{section}{References}
\bibliographystyle{JHEP}
\bibliography{main,main-phi}
\clearpage

\end{document}